\documentclass[pre, twocolumn, showpacs,amsmath,amssymb]{revtex4}

\usepackage{hyperref}
\usepackage{graphicx}
\usepackage{dcolumn}
\usepackage{xcolor}
\usepackage{braket}
\usepackage{commath}
\usepackage{adjustbox}
\usepackage{comment} 
\usepackage{ulem}
\usepackage{tikz}
\usetikzlibrary{shapes}

\newcommand{\bs}[1]{{\boldsymbol{#1}}}
\newcommand{\bk}{\bs{k}}
\newcommand{\bq}{\bs{q}}

\newcommand{\br}{\bs{r}}
\newcommand{\be}{\bs{\epsilon}}

\begin{document}

\title{Coherent multiple scattering of light in (2+1) dimensions}

\author{Nicolas Cherroret}
\email{cherroret@lkb.upmc.fr}
\affiliation{Laboratoire Kastler Brossel, Sorbonne Universit\'e, CNRS, ENS-PSL University, Coll\`ege de France; 4 Place Jussieu, 75005 Paris, France}


\begin{abstract}
We formulate a multiple scattering theory of light in media spatially disordered along two directions and homogeneous along the third one,
without making any paraxial approximation on the wave equation and fully 
treating
the vector character of light.
With this formalism, we calculate the distribution of transverse momenta of a beam as it evolves along the optical axis, and unveil a phenomenon not captured by the paraxial equation: a cross-over from a scalar to a vector regime, visible in the coherent backscattering peak as polarization gets randomized.
\end{abstract}


\maketitle

\section{Introduction}

When light propagates mostly along a given direction $z$ in a weakly heterogeneous medium, its evolution is in first approximation governed by the paraxial wave equation. 
The latter has the same structure as the time-dependent Schr\"odinger equation, the longitudinal coordinate $z$ playing the role of time \cite{Agrawal95, Rosanov02}. 
In Kerr media this idea was recently applied to study optical analogues of quantum matter-wave phenomena, like Bose condensation \cite{Sun12, Santic18} or superfluidity \cite{Wan07, Michel17}.
 In a similar spirit, in spatially disordered materials paraxial light propagation was exploited to observe the coherent backscattering (CBS) effect without interface \cite{Boguslawski17}, in a configuration that reproduces cold-atomic setups \cite{Cherroret12, Jendrzejewski12}. In the same context, transverse Anderson localization of optical wave packets \cite{Raedt89} was reported in the paraxial limit \cite{Schwartz07, Boguslawski13}. 
A major drawback  of the paraxial equation, however, is that it neglects the vector nature of light. 
In presence of disorder,
how polarization effects manifest themselves beyond the paraxial approximation and when the latter breaks down is not well known.  
Clarifying these questions is essential to clearly identify the regime where a scalar description of light can be used in experiments that operate close to the paraxial limit.
More generally, understanding the role of the vector nature of light in heterogeneous media is required for a proper characterization of disorder-induced interference effects like CBS, known to be sensitive to polarization in general \cite{Stephen86, Wolf88}. The question is even crucial for Anderson localization of light, which was shown  to be severely altered in random ensembles of scatterers when vector effects are accounted for \cite{Skipetrov14, Bellando14, Maximo15}.

\begin{figure}
\centering
\includegraphics[width=0.87\linewidth]{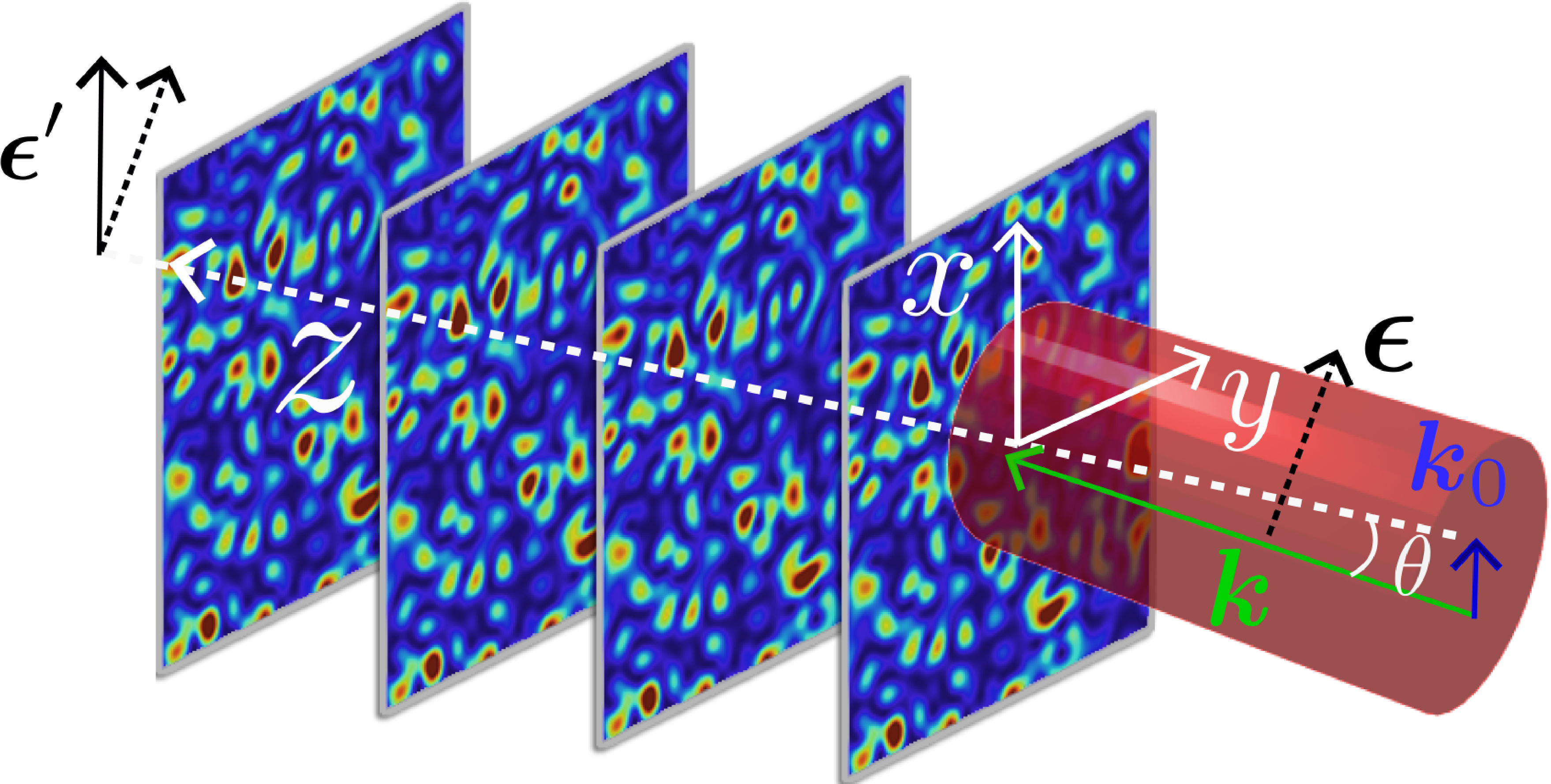}
\caption{
\label{3Dscheme}
We consider the propagation of a quasi plane-wave beam of transverse wave vector $\bk_0$ and polarization $\be$ through a medium spatially disordered along $x$ and $y$ and homogeneous along the optical axis $z$. Light is detected on a polarization channel $\be'$ belonging to the plane $(x,y)$.
}
\end{figure}

In this paper, we develop a general theory of multiple scattering of light in media disordered  along two transverse directions $x$ and $y$ and homogeneous along the third one $z$ (the optical axis). This theory does not rely on the paraxial approximation, but is constructed from the exact Helmholtz wave equation. As a concrete example, we calculate the disorder-average distribution $\overline{|\textbf{E}(\bk_\perp,z)|^2}$ of transverse momenta of an incident plane-wave beam of transverse momentum $\boldsymbol{k}_0$  as it propagates along $z$, see Fig. \ref{3Dscheme}.
In this configuration, it was theoretically shown in the framework of the paraxial (or Schr\"odinger) equation that this distribution evolves toward a pedestal ring of radius $|\boldsymbol{k}_0|$ due to the diffusive randomization of momenta, on top of which a CBS peak arises around $\bk_\perp=-\bk_0$ \cite{Cherroret12, Ghosh15}. This phenomenon was recently observed experimentally in photorefractive materials \cite{Boguslawski17}.
By revisiting it  within a full vector treatment, we discover an interesting effect beyond the paraxial approximation:
the momentum distribution 
crosses-over from a scalar to a vector regime as $z$ exceeds a characteristic scale $z_{p}$, 
identified as the time over which the polarization direction is randomized. As long as $z\ll z_{p}$, the polarization remains fixed, the paraxial equation applies and the physics of CBS, in particular, is the one of a scalar wave.
In strong contrast, when $z\gg z_{p}$ the paraxial theory breaks down and the CBS peak starts to non-trivially depend on polarization. In particular, for a circularly-polarized incident beam the CBS peak is visible only from light of opposite circular polarization, as imposed by the reciprocity principle. 
We also find that the scale $z_p$ strongly depends on the injection  angle of the incident beam ($\theta$ in Fig. \ref{3Dscheme}), thus offering a convenient way of switching from a scalar to a vector regime for light in a 2D disordered environment.

\section{Optical momentum distribution}
Our starting point is the Helmholtz equation for the complex electric field $\textbf{E}(\br)=\textbf{E}(x,y,z)$:
\begin{equation}
\left[\Delta\delta_{ij}-\nabla_i\nabla_j+k^2\delta_{ij}-V(\br_\perp)\delta_{ij}
\right]E_j(\boldsymbol{r})=0,
\label{Helmholtz_eq}
\end{equation}
where $k^2=\omega^2\bar{\mathcal{\epsilon}}/c^2$, with $\omega$ the light carrier frequency.
The disorder ``potential'' $V(\br_\perp)=-k^2\delta\epsilon(\br_\perp)/\bar{\epsilon}$ stems from spatial fluctuations $\delta\epsilon(\br_\perp)$ of the refractive index around an average value $\bar{\epsilon}$. Disorder is only along $\br_\perp=(x,y)$, not along $z$. In Eq. (\ref{Helmholtz_eq}) and in the following, summation over repeated indices $j=x,y,z$ is implied. For simplicity, we choose the disorder to be Gaussian distributed and uncorrelated of strength $\gamma$:
\begin{equation}
\overline{V(\br_\perp)V(\br_\perp')}\equiv k^4\frac{\overline{\delta\epsilon(\br_\perp)\delta\epsilon(\br_\perp')}}{\bar{\epsilon}^2}=k^4\gamma\delta(\br_\perp-\br_\perp'),
\label{deltacor_eq}
\end{equation}
with $\overline{\delta\epsilon(\br_\perp)}=0$.
The main goal of the paper is to evaluate the disorder-average distribution $\overline{|\be'\cdot\textbf{E}(\bk_\perp,z)|^2}$ of transverse momenta as a function of the longitudinal coordinate $z$. In this definition, the output light beam, at $z$, is detected in the direction $\bk_\perp$ and along a polarization axis $\be'$ that lies in the plane $(x,y)$, see Fig. \ref{3Dscheme}. 
 We model the incident beam at $z=0$, $E_i(\bk_\perp,z\!=0)=A(\bk_\perp)\epsilon_i$, by a field distribution $A(\bk_\perp)$ peaked around $\bk_\perp=\bk_0$ and polarized along the complex unit vector $\be$ perpendicular to $\bk$  (Fig. \ref{3Dscheme}). The field at $z$ is given by:
\begin{align}
E_j(\bk_\perp,z)=
 2ik \int\frac{d^2\bk_\perp'}{(2\pi)^2}
\langle\bk_\perp |G_{ij}(z)|\bk_\perp'\rangle
A(\bk_\perp')\epsilon_i,
\label{EG_eq0}
\end{align}
where $G_{ij}$ is the Green tensor of the Helmholtz equation (\ref{Helmholtz_eq}). The tensor $2ikG_{ij}(z)$ can be interpreted as the medium's transmission coefficient from $z=0$ to $z$ \cite{Berkovits94}. 
In order to obtain the momentum distribution, we average the square modulus of Eq. (\ref{EG_eq0}) contracted with $\be'$,
introducing the Fourier transform $G_{ij}(z)\equiv \int_{-\infty}^\infty dk_z/(2\pi)G_{ij}(k_z) e^{i k_z z}$. This gives \cite{Cherroret12}:
\begin{align}
&\overline{|\be'\cdot\textbf{E}(\bk_\perp,z)|^2}=4k^2\int_0^\infty\!\frac{dk_z}{2\pi}\int _{-\infty}^\infty\!\frac{dq_z}{2\pi}\int\!\frac{d^2\bk_\perp'}{(2\pi)^2}
|A(\bk'_\perp)|^2\nonumber\\
&\times
\epsilon_i^*\epsilon_j'\overline{\langle\bk_\perp |G_{ij}(k_z^+)|\bk_\perp'\rangle \langle\bk_\perp' |G^\dagger_{kl}(k_z^-)|\bk_\perp\rangle }
\epsilon_k^{*\prime}\epsilon_l\, e^{i q_z z},
\label{EG_eq}
\end{align}
where $k_z^\pm\equiv k_z\pm q_z/2$ and we resorted to  translation invariance in plane $(x,y)$ to remove one integral over $\bk_\perp'$. We now assume a normalized, incident plane wave, $|A(\bk_\perp)|^2\simeq (2\pi)^2\delta(\bk_\perp-\bk_0)$, and decompose the average product of two Green tensors in a standard way 
\cite{Sheng95}:
\begin{align}
&\overline{|\be'\cdot\textbf{E}(\bk_\perp,z)|^2}=4k^2\epsilon_i^*\epsilon_j'\epsilon_k^{*\prime}\epsilon_l \times
\nonumber\\
&\int_0^\infty\frac{dk_z}{2\pi}
\overline{G}_{i\alpha}(\bk_0,k_z)
\overline{G}^*_{k\beta}(\bk_0,k_z)
\overline{G}_{j\gamma}(\bk_\perp,k_z)
\overline{G}_{l\delta}^*(\bk_\perp,k_z)\nonumber\\
&\times\int_{-\infty}^\infty \frac{dq_z}{2\pi}\langle\bk_\perp,\bk_0|\Gamma_{\alpha\beta,\gamma\delta}(k_z,q_z)|
\bk_0,\bk_\perp\rangle e^{i q_z z}.
\label{EG_eq_simplified}
\end{align}
In writing Eq. (\ref{EG_eq_simplified}), we introduced four average Green tensors defined through the relation $\overline{\langle\bk_\perp |G_{ij}(k_z)|\bk_\perp'\rangle}=\delta(\bk_\perp-\bk'_\perp)\overline{G}_{ij}(\bk_\perp,k_z)$ implied by translation invariance on average in the plane $(x,y)$. 
We also dropped the $q_z$ dependence of these tensors, which is justified in the large-$z$ limit \cite{Montambaux07} (the so-called diffusive regime, see below for the precise condition).
Physically, the first two Green tensors describe the average propagation of the incoming beam at $\bk_0$ to the first scattering event in the material, and the last two ones the average propagation of the outgoing beam at $\bk_\perp$ from the last scattering event to the detection point. The structure tensor $\boldsymbol{\Gamma}$, on the other hand, contains all information about the multiple scattering process from the first to the last scattering event. Evaluating the momentum distribution requires the computation of the tensors $\overline{\textbf{G}}$ and $\boldsymbol{\Gamma}$. We execute this program in the next two sections.

\section{Average Green tensor in (2+1) dimensions}

We first examine the average Green tensor $\overline{\textbf{G}}(\bk_\perp,k_z)$, given by the vector Dyson equation \cite{Stephen86, vanTiggelen96, Busch05}
\begin{align}
\overline{\textbf{G}}(\bk_\perp,k_z)
=\left[\textbf{G}^{(0)}(\bk_\perp,k_z)^{-1}
-\boldsymbol{\Sigma}(\bk_\perp,k_z)\right]^{-1},
\label{Dyson_eq}
\end{align}
where $\textbf{G}^{(0)}$ is the free-space, transverse Green tensor (here and in the following, we neglect near-field effects): 
\begin{align}
G^{(0)}_{ij}(\bk_\perp,k_z)=\frac{\delta_{ij}-\hat{k}_i\hat{k}_j}{k^2-\bk_\perp^2-k_z^2+i0^+},
\end{align}
where $\bk\equiv(\bk_\perp,k_z)$.
In this paper we assume a weakly disordered material, so that the Born approximation can be used for calculating the self-energy tensor $\Sigma_{ij}$.
For a delta-correlated potential,  Eq. (\ref{deltacor_eq}), this gives \cite{Montambaux07}
\begin{equation}
\Sigma_{ij}(k_z)=\int \frac{d^2\bk_\perp'}{(2\pi)^2}\frac{\delta_{ij}-\hat{k}'_i\hat{k}'_j}{k^2-\bk_\perp'^2-k_z^2+i0^+}.
\label{Sigma_def_eq}
\end{equation}
Let us comment on the physical meaning of the longitudinal wave vector $k_z$. At weak disorder, the Green tensors $\overline{\textbf{G}}(\bk_0,k_z)$ in the general formulation (\ref{EG_eq_simplified}) of the momentum distribution are peaked around $k_z^2=k^2-k_0^2$. Therefore, the parameter $\hat{k}_z\equiv k_z/k\simeq 1-k_0/k$, between 0 and 1, quantifies the deviations of the incident beam from paraxiality. 
$\Sigma_{ij}(k_z)$ and $G_{ij}(\bk_\perp,k_z)$ can be found for any value of $k_z$ by exact calculation of the integral in Eq. (\ref{Sigma_def_eq}) and diagonalization of Eq. (\ref{Dyson_eq}). 
We give their expressions in the appendix as they are rather cumbersome. They will be needed for the proper evaluation of the structure factor presented in Sec. \ref{structure_factor_sec}.
For the present discussion, however, let us focus our attention on the value of $\overline{\textbf{G}}$ for $\hat{k}_z\simeq 1$. In this limit, the imaginary part of the self-energy $\text{Im}\Sigma_{ij}(k_z)\simeq-\delta_{ij}\gamma k^4/4$ and:
\begin{equation}
\overline{G}_{ij}(\bk_\perp,k_z)\simeq \frac{\delta_{ij}-\hat{k}_i\hat{k}_j}{k^2-\bk_\perp^2-k_z^2+ik/z_s},
\label{av_E_paraxial}
\end{equation}
where $z_s\equiv 4/(\gamma k^3)$. 
By Fourier transforming Eq. (\ref{av_E_paraxial}) with respect to $k_z$ and inserting the result into Eq. (\ref{EG_eq0}), we infer:
\begin{align}
\overline{\textbf{E}}(\bk_\perp,z)\simeq \textbf{E}(\bk_\perp,z=0)e^{ikz-z/2z_s},
\end{align}
which  describes the usual depletion of the incident mode due to scattering as it evolves along $z$ \cite{Montambaux07}, at a rate governed by the effective \textit{scattering time} $z_s$.
Eq. (\ref{av_E_paraxial}) also provides the regime of validity of the Born approximation: $|k^2-k_z^2|\sim k_0^2\ll k/z_s$. This weak-disorder condition can be rewritten as the familiar $k_0\ell \gg1$ \cite{Montambaux07}, where $\ell\equiv\hat{k}_0 z_s$ is the mean free path, proportional to the effective transverse velocity $\hat{k}_0$ of the incident beam.

Let us remind that Eq. (\ref{av_E_paraxial}) was obtained assuming $\hat{k}_z\simeq1$, i.e. quasi-paraxiality. This can be confirmed a posteriori by noticing that Eq. (\ref{av_E_paraxial}) indeed coincides with the solution of the paraxial wave equation. 
The latter is traditionally obtained by writing $\textbf{E}(\br)=\boldsymbol{\mathcal{E}}(\br)e^{ikz}$ and assuming slow variations of the envelope along $z$, $|\partial^2\mathcal{E}/\partial z^2|\ll k|\partial\mathcal{E}/\partial z|$, and of the permittivity in the plane $(x,y)$, $|\nabla_\perp \epsilon|\ll k$ \cite{Agrawal95, Rosanov02}. Under these conditions, the envelope $\boldsymbol{\mathcal{E}}$ is mostly transverse and the Helmholtz equation (\ref{Helmholtz_eq}) simplifies
\begin{equation}
\left[\Delta\delta_{ij}+2ik\delta_{ij}\frac{\partial}{\partial z}-V(\br_\perp)\delta_{ij}
\right]\mathcal{E}_j(\boldsymbol{r})=0,
\label{Paraxial_eq}
\end{equation}
which mimicks a time-dependent Schr\"odinger equation where $z$ plays the role of time. By applying to Eq. (\ref{Paraxial_eq}) the scalar version of the Green function technique introduced above \cite{Cherroret12}, we recover, in the Born approximation, that $\overline{\boldsymbol{\mathcal{E}}}(\bk_\perp,z)=\boldsymbol{\mathcal{E}}(\bk_\perp,z=0)\exp(-z/2z_s)$.


\section{Structure factor in (2+1) dimensions}
\label{structure_factor_sec}

\subsection{Classical contribution}
\label{classical_sec}

\begin{figure}
\centering
\includegraphics[width=1\linewidth]{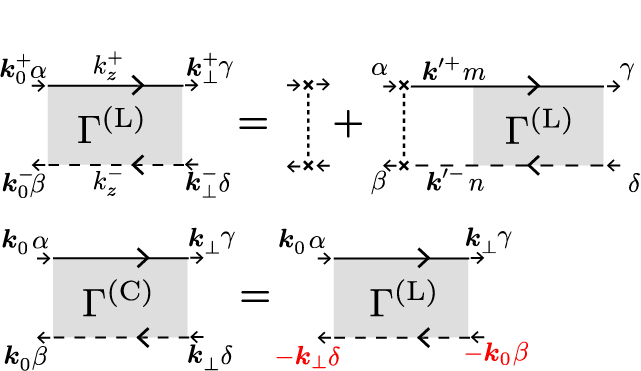}
\caption{
\label{LCLC}
Top: diagrammatic representation of the Bethe-Salpether equation (\ref{iterative_ladder}) for the ladder series $\Gamma^\text{(L)}$. Upper solid lines symbolize the Green tensor $\textbf{G}$, and lower dashed lines its complex conjugate. Vertical dotted lines refer to the correlation function of the disorder potential, Eq. (\ref{deltacor_eq}).
Bottom: reciprocity relation (\ref{reciprocity_eq}) between ladder and crossed series.
}
\end{figure}

Under the weak-disorder condition $k_0\ell\gg1$, the structure factor has two well-known contributions: the series of ladder and of crossed diagrams, which respectively describe  classical diffusion and coherent backscattering. 
We first focus on the ladder series. It gives a contribution $\boldsymbol{\Gamma}^\text{(L)}$ to $\boldsymbol{\Gamma}$, which obeys the Bethe-Salpether equation \cite{Stephen86, Sheng95, Montambaux07, vanTiggelen96, Busch05}:
\begin{align}
&\Gamma^\text{\text{(L)}}_{\alpha\beta,\gamma\delta}(k_z,q_z, \bq)=\gamma k^4 \delta_{\alpha\gamma}\delta_{\beta\delta}+
\gamma k^4\Gamma^\text{(L)}_{mn,\gamma\delta}(k_z,q_z, \bq)\nonumber\\
&\times\int \frac{d^2\bk'}{(2\pi)^2}
\overline{G}_{\alpha m}(\bk'^+\!,k_z^+)
\overline{G}^*_{\beta n}(\bk^{\prime-}\!,k_z^-),
\label{iterative_ladder}
\end{align} 
where $\bk'^\pm=\bk'\pm\bq/2$ and we introduced the short-hand notation
\begin{eqnarray}
\Gamma^\text{(L)}_{\alpha\beta,\gamma\delta}(k_z,q_z,\bq)\!\equiv\!
\langle
\bk_\perp^+,\bk_0^-
|\Gamma^\text{(L)}_{\alpha\beta,\gamma\delta}(k_z,q_z)|
\bk_0^+,\bk_\perp^-
\rangle.
\label{GammaL_def}
\end{eqnarray}
The Bethe-Salpether equation is shown digrammatically  in Fig. \ref{LCLC}, which also indicates the conventions used for wave vectors and polarization indices.
Notice that the 
object (\ref{GammaL_def}) is slightly more general than the structure factor 
in Eq. (\ref{EG_eq}), because it involves incoming and outgoing wave vectors $\bk_0^\pm=\bk_0\pm\bq/2$ and $\bk_\perp^\pm=\bk_\perp\pm\bq/2$ instead of simply $\bk_0$ and $\bk_\perp$. The additional variable $\bq$ will be useful later for evaluating the crossed contribution.
The definition (\ref{GammaL_def}) moreover expresses that $\Gamma^\text{(L)}_{\alpha\beta,\gamma\delta}(k_z,q_z,\bq)$ does not 
explicitly depend on 
$\bk_0$ and $\bk_\perp$ [this is why we could take $\Gamma^\text{(L)}$ out of the integral over $\bk'$ in Eq. (\ref{iterative_ladder})]. Such property is a consequence of our  model of  uncorrelated disorder, Eq. (\ref{deltacor_eq}), which scatters light isotropically. 

The momentum distribution (\ref{EG_eq_simplified}) involves the ladder structure factor at $\bq=0$. To find it,
we set $\bq=0$ in Eq. (\ref{iterative_ladder}) and  introduce the tensor
\begin{equation}
Q_{\alpha\beta,\gamma\delta}(k_z,q_z)\!\equiv\!\gamma k^4\!\!
\int\! \frac{d^2\bk'}{(2\pi)^2}
\overline{G}_{\alpha \gamma}(\bk'\!,k_z^+)
\overline{G}^*_{\beta \delta}(\bk^{\prime}\!,k_z^-)
\label{Qdef}
\end{equation}
as well as $I_{\alpha\beta,\gamma\delta}\equiv\gamma k^4\delta_{\alpha\gamma}\delta_{\beta\delta}$. Eq. (\ref{iterative_ladder}) thus reads
\begin{align}
\boldsymbol{\Gamma}^\text{(L)}= \textbf{I}+\textbf{Q}\cdot\boldsymbol{\Gamma}^\text{(L)} =\gamma k^4\sum_{m=0}^\infty \textbf{Q}^m.
\label{gammasum}
\end{align}
The infinite sum can be carried out by diagonalizing $\textbf{Q}$ in terms of orthogonal projectors $\boldsymbol{\Pi}^{(n)}$ and associated eigenvalues $\lambda_n$ \cite{Muller02}:
\begin{align}
\textbf{Q}=\sum_n \lambda_n\boldsymbol{\Pi}^{(n)},
\label{Q_decompo}
\end{align}
where $\boldsymbol{\Pi}^{(n)}\!\cdot\!\boldsymbol{\Pi}^{(n')}\!\!=\!\delta_{nn'}$ and $\sum_n\!\!\boldsymbol{\Pi}^{(n)}\!\!=\!\textbf{1}$. Substituting the formal decomposition (\ref{Q_decompo}) for $\textbf{Q}$ in Eq. (\ref{gammasum}), we obtain:
\begin{align}
\boldsymbol{\Gamma}^\text{(L)}=\sum_{n} \frac{\gamma k^4}{1-\lambda_n}\boldsymbol{\Pi}^{(n)}.
\label{gammasummed}
\end{align}
The problem thus reduces to finding the eigenmodes $(\lambda_n,\boldsymbol{\Pi}^{(n)})$. 
This task requires to use the complete expression of the average Green tensor, Eq. (\ref{AvG_final_eq}) of the appendix. Due to the complex anisotropic structure of the latter, the diagonalization of the tensor $\textbf{Q}$ is considerably more difficult than in usual 3D isotropic disordered media, which are rotationally invariant on average \cite{Stephen86}.

We have been able to derive the modes $(\lambda_n,\boldsymbol{\Pi}^{(n)})$ in the diffusive regime $z\gg z_s$, equivalently $q_z z_s\ll1$, by unfolding the coordinates of $Q_{\alpha\beta,\gamma\delta}$ onto a $9\times9$ matrix to diagonalize. 
The polarization space of this matrix can be decomposed into $6$ eigensubspaces, three of them associated with a twice-degenerate eigenvalue. The exact expressions of the $(\lambda_n,\boldsymbol{\Pi}^{(n)})$ are rather cumbersome so they are reported in the appendix for clarity.
\begin{figure}
\centering
\includegraphics[width=0.95\linewidth]{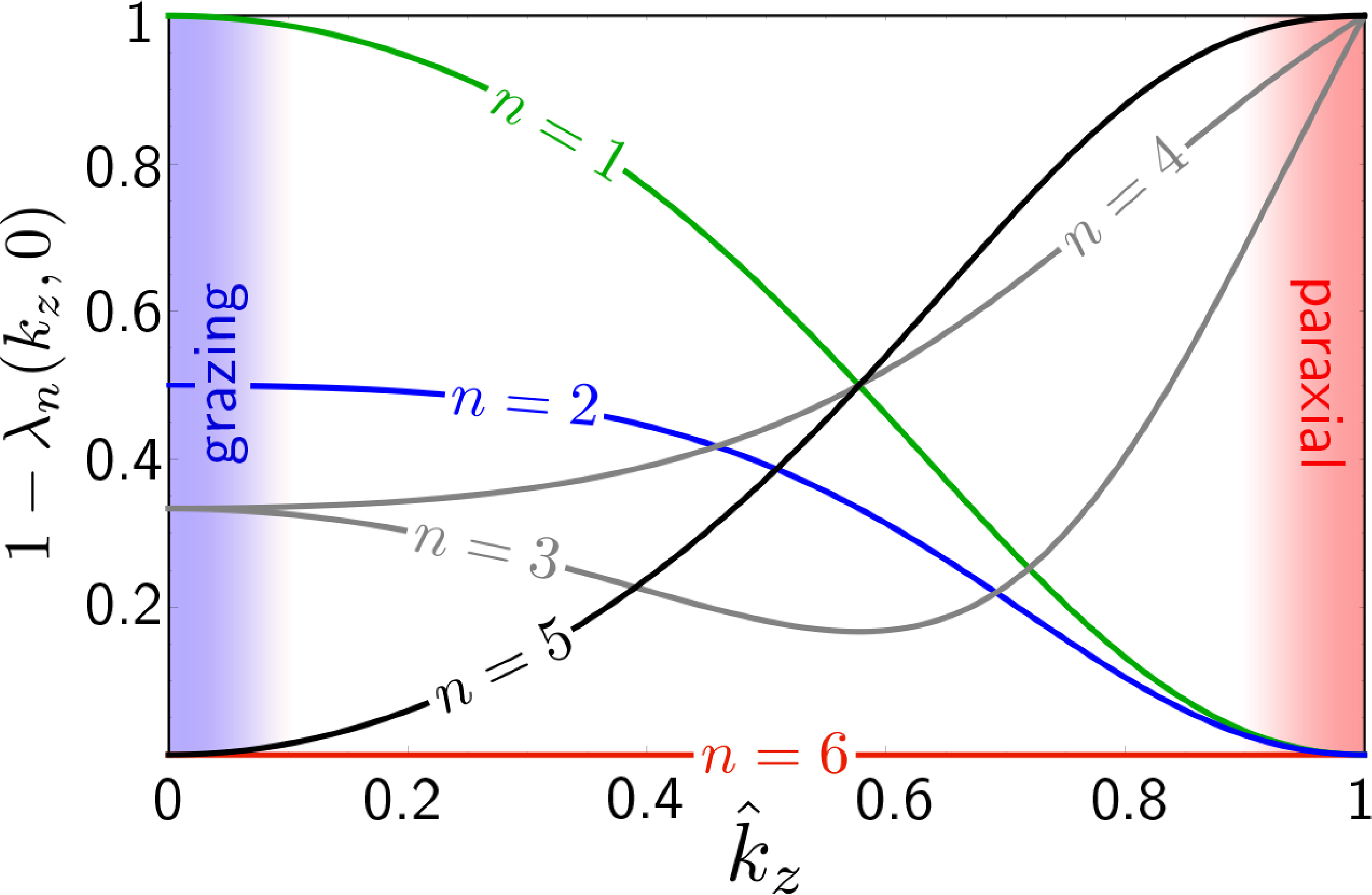}
\caption{
\label{lifetimes_fig}
Factors $1-\lambda_n(k_z,q_z=0)$ of the mode decomposition (\ref{gammasummed}).
The mode $n=6$ fulfills $1-\lambda_6(k_z,0)=0$ for all $\hat{k}_z$. It thus always controls the large $z$ (small $q_z$) limit of the structure factor (\ref{gammaLsum}).
In the experimentally-relevant limit where $\hat{k}_z$ is close to 1, the two modes $n=1$ and $2$ obey $1-\lambda_{1,2}(k_z,0)\ll 1$, so they also contribute in general. The modes $n=3,4$ and $5$ are, on the other hand, strongly attenuated in this regime.
}
\end{figure}
The eigenvalues are of the form $\lambda_n(k_z,q_z)=\lambda_n(k_z,0)-i q_z \mu_n(k_z)$
, such that 
\begin{eqnarray}
\Gamma_{\alpha\beta,\gamma\delta}^\text{(L)}(k_z,q_z,0)=\sum_{n=1}^6
\frac{\gamma k^4{\Pi}^{(n)}_{\alpha\beta,\gamma\delta}(k_z)}{1-\lambda_n(k_z,0)+i q_z \mu_n(k_z)}.
\label{gammaLsum}
\end{eqnarray}
In the diffusive regime $q_z z_s\ll1$, Eq. (\ref{gammaLsum}) is dominated by the smallest factors $1-\lambda_n(k_z,0)$. These quantities are displayed in Fig. \ref{lifetimes_fig} as a function of $\hat{k}_z=k_z/k$, from the paraxial regime ($\hat{k}_z\simeq 1$) to the limit of grazing incidence ($\hat{k}_z\ll 1$). 
The plots reveal two remarkable properties. 
First, the mode $n=6$ satisfies $1-\lambda_6(k_z,0)=0$
whatever $\hat{k}_z$. This implies a diffusive behavior at large distances, $\Gamma^\text{(L)}(k_z,q_z,0)\propto 1/(iq_z)$, which is needed to guarantee the conservation of normalization \cite{Barabanenkov95a}. It is worth noticing that keeping the complete expression (\ref{AvG_final_eq}) of $\overline{G}$ is absolutely \textit{crucial} to recover this property.
Second, in the vicinity of the paraxial regime, $\hat{k}_z\simeq 1$, two additional modes, $n=1$ and $2$, have very 
small denominators $1-\lambda_{1,2}(k_z,0)$ and
therefore persist up to very long times. This phenomenon, which constitutes the main finding of the paper,  signals a cross-over in light propagation, occuring around a characteristic scale $z_p$ that will be discussed in detail in Sec. \ref{sec_results}. 
In short, for $z\gg z_p$ the modes $n=1$ and $2$ are negligible, whereas for $z\ll z_p$ they must be accounted for, which qualitatively modifies the properties of the structure factor.
In the strict limit $\hat{k}_z= 1$ (beam aligned along the optical axis), we even have $1-\lambda_6(k_z,0)=1-\lambda_1(k_z,0)=1-\lambda_2(k_z,0)=0$, so that the three modes $n=1,2$ and $6$ equally contribute to Eq. (\ref{gammaLsum}) up to arbitrary large $z$.

\subsection{CBS contribution}
 
The CBS interference peak is described by the series of crossed diagrams denoted by $\Gamma^\text{(C)}$. By virtue of reciprocity, $\Gamma^\text{(C)}$ is related to $\Gamma^\text{(L)}$ through \cite{Montambaux07}
\begin{align}
&\langle\bk_\perp,\bk_0|\Gamma^\text{(C)}_{\alpha\beta,\gamma\delta}(k_z,q_z)|
\bk_0,\bk_\perp\rangle=\nonumber\\
&\hspace*{1.2cm}
\langle\bk_\perp,-\bk_\perp|\Gamma^\text{(L)}_{\alpha\delta,\gamma\beta}(k_z,q_z)|
\bk_0,-\bk_0\rangle
\label{reciprocity_eq}
\end{align}
as illustrated diagrammatically in Fig. \ref{LCLC}. According to Eq. (\ref{GammaL_def}), we thus obtain the CBS contribution from $\Gamma^\text{(L)}_{\alpha\beta,\gamma\delta}(k_z,q_z, \bq)$, by interchanging $\beta$ and $\delta$ and setting $\bq=\bk_\perp+\bk_0$. This requires to solve Eq. (\ref{iterative_ladder}) for $\bq\ne 0$. 
We have accomplished this task in the diffusive regime $|\bq|z_s\ll1$. This leads to an extra dependence of the $\lambda_n$ on $\bq^2$
(the $\bq^2$ corrections to the projectors $\boldsymbol{\Pi}^{(n)}$ are, on the other hand, negligible):
\begin{eqnarray}
&&\Gamma_{\alpha\beta,\gamma\delta}^\text{(C)}(k_z,q_z)=\Gamma_{\alpha\delta,\gamma\beta}^\text{(L)}(k_z,q_z,\bk_0+\bk_\perp)=\label{gammaCsum}\\
&&\sum_{n=1}^6\frac{\gamma k^4\Pi^{(n)}_{\alpha\delta,\gamma\beta}(k_z)}{1-\lambda_n(k_z,0)+i q_z \mu_n(k_z)+\nu_n(k_z)(\bk_\perp+\bk_0)^2}.\nonumber
\end{eqnarray}
The explicit expression of the new factors $\nu_n(k_z)$ is given in the appendix for the three modes $n=1,2$ and $6$, relevant at large $z$ near the paraxial limit.

\section{Results}
\label{sec_results}

\subsection{The paraxial regime revisited}

We now derive the explicit expression of the momentum distribution (\ref{EG_eq_simplified}) for a typical setup where the incident beam is almost aligned with the optical axis ($\hat{k}_0\ll 1$). In this configuration, it is sufficient to use Eq. (\ref{av_E_paraxial}) to evaluate the four average Green tensors in Eq. (\ref{EG_eq_simplified}). The first two ones are strongly peaked around $k_z^2\simeq k^2-k_0^2$, which allows us to approximate $\Gamma_{\alpha\beta,\gamma\delta}(k_z,q_z)\simeq\Gamma_{\alpha\beta,\gamma\delta}(\sqrt{k^2-k_0^2},q_z)\simeq\Gamma_{\alpha\beta,\gamma\delta}(k,q_z)$.
The integral over $k_z$ then applies to the product of four Green tensors only
\begin{align}
&\int_0^\infty\frac{dk_z}{2\pi}
\overline{G}_{i\alpha}(\bk_0,k_z)
\overline{G}^*_{k\beta}(\bk_0,k_z)
\overline{G}_{j\gamma}(\bk_\perp,k_z)
\overline{G}_{l\delta}^*(\bk_\perp,k_z)\nonumber\\
&=\frac{z_s/2k^2}{(2k/z_s)^2+(\bk_\perp^2-\bk_0^2)^2}
\delta_{i\alpha}\delta_{k\beta}\delta_{j\gamma}\delta_{l\delta}.
\label{kz_integral_eq}
\end{align}
We dropped here the tensorial parts of the type $\hat{k}_i\hat{k}_\alpha$ of the Green tensors, which are negligible once contracted with the incoming and outgoing polarization vectors $\be$ and $\be'$.
Let us now deal with the structure factor in Eq. (\ref{EG_eq_simplified}). As seen in Sec. \ref{classical_sec}, when $\hat{k}_0\ll 1$ (corresponding to $\hat{k}_z\simeq1$) only the three modes $n=1,2$ and $6$ contribute to $\boldsymbol{\Gamma}$. 
Eqs. (\ref{gammaLsum}) and (\ref{gammaCsum}) then lead to:
\begin{align}
&\int_{-\infty}^\infty \frac{dq_z}{2\pi}\langle\bk_\perp,\bk_0|\Gamma_{\alpha\beta,\gamma\delta}(k_z,q_z)|
\bk_0,\bk_\perp\rangle e^{i q_z z}\simeq
\label{qz_integral_eq}\\
&\frac{(\gamma k^4)^2}{4k}\sum_{n=1,2,6}\!
e^{-z_n/z_s}
\left[\Pi^{(n)}_{\alpha\beta,\gamma\delta}+\Pi^{(n)}_{\alpha\delta,\gamma\beta}e^{-D_n(\bk_\perp+\bk_0)^2z}\right].\nonumber
\end{align}
The first term inside the brackets is the ladder contribution, while the second describes the CBS peak.
The projectors $\boldsymbol{\Pi}^{(n)}$ and their corresponding lifetimes $z_n\equiv \mu_n(k_z)/[1-\lambda_n(k_z,0)]$ take a particularly simple form in the regime $\hat{k}_z\simeq 1$ considered here:
\begin{align}
&\Pi^{(6)}_{\alpha\beta,\gamma\delta}\simeq\frac{1}{2}\delta_{\alpha\beta}\delta_{\gamma\delta},\ \ z_6=\infty\nonumber\\
&\Pi^{(1)}_{\alpha\beta,\gamma\delta}\simeq\frac{1}{2}(\delta_{\alpha\gamma}\delta_{\beta\delta}-\delta_{\alpha\delta}\delta_{\beta\gamma}),\ \ z_1\simeq\frac{4z_s}{3\hat{k}_0^4}
\label{Pi_expressions_eq}\\
&\Pi_{\alpha\beta,\gamma\delta}^{(2)}\simeq\frac{1}{2}(\delta_{\alpha\gamma}\delta_{\beta\delta}+\delta_{\alpha\delta}\delta_{\beta\gamma}-\delta_{\alpha\beta}\delta_{\gamma\delta}),\ \ z_2\simeq\frac{8z_s}{5\hat{k}_0^4}.\nonumber
\end{align}
The quantities $D_n=\nu_n(k_z)/\mu_n(k_z)$ in Eq. (\ref{qz_integral_eq}) are the diffusion coefficients of each mode.
When $\hat{k}_z\simeq 1$, they all equal (see appendix)
\begin{equation}
D_n\simeq\frac{\hat{k}_0^2 z_s}{2}\equiv D.
\label{diff_eq}
\end{equation}
In this expression, we recall that $z_s$ is the effective mean free time and $\hat{k}_0$ is the transverse velocity of the incident beam. The factor 2, finally, refers to the dimensionality of the disorder.
Making use of Eqs. (\ref{kz_integral_eq}), (\ref{qz_integral_eq}), (\ref{Pi_expressions_eq}) and (\ref{diff_eq}), we rewrite Eq. (\ref{EG_eq_simplified}) as:
\begin{align}
\overline{|\be'\cdot\textbf{E}(\bk_\perp,z)|^2}=\frac{8k}{z_s}\frac{\mathcal{F}(\be,\be',z,\bk_\perp)}{(2k/z_s)^2+(\bk_\perp^2-\bk_0^2)^2},
\label{momentum_final}
\end{align}
where the function $\mathcal{F}=\mathcal{F}_\text{L}+\mathcal{F}_\text{C}$ splits into a diffusive contribution,
\begin{align}
&\mathcal{F}_\text{L}(\be,\be',z)=\frac{1}{2}\left[1+(|\be\cdot\be'^*|^2-|\be\cdot\be'|^2)e^{-z/z_1}\right.\nonumber\\
&\left.+(|\be\cdot\be'^*|^2+|\be\cdot\be'|^2-1)e^{-z/z_2}\right],
\label{FL_eq}
\end{align}
and a contribution of the coherent backscattering peak:
\begin{align}
&\mathcal{F}_\text{C}(\be,\be',z,\bk_\perp)=\frac{1}{2}\left[|\be\cdot\be'|^2
+(|\be\cdot\be'^*|^2-1)e^{-z/z_1}\right.\nonumber\\
&\left.+(|\be\cdot\be'^*|^2-|\be\cdot\be'|^2+1)e^{-z/z_2}\right]e^{-D(\bk_\perp+\bk_0)^2z}.
\label{FC_eq}
\end{align}
Eqs. (\ref{momentum_final}, \ref{FL_eq}, \ref{FC_eq}) for the momentum distribution constitute the main result of the paper. The distribution (\ref{momentum_final}) is sketched in the upper-right inset of Fig. \ref{F_plots_fig} in the plane $(k_x,k_y)$.
It consists of a diffusive ring of radial width $1/z_s$, on top of which the CBS peak stands around $\bk_\perp=-\bk_0$  \cite{Cherroret12, Boguslawski17}.
This peak has a narrow width $1/\sqrt{Dz}$ that decreases with $z$ at a rate controlled by the diffusion coefficient $D$ \cite{Tourin97, Jendrzejewski12, Cobus16, Cherroret12}.

\subsection{Cross-over from a scalar to a vector regime}

We now come to the central result of the paper:
Eqs. (\ref{FL_eq}) and (\ref{FC_eq}) highlight a cross-over between two well distinct regimes. When $z\ll z_{1,2}\sim z_s/\hat{k}_0^4$ first, we have
\begin{equation}
\mathcal{F}(\be,\be',z\ll z_{1,2},\bk_\perp)=|\be\cdot\be'^*|^2\left[1+e^{-D(\bk_\perp+\bk_0)^2z}\right],
\end{equation}
corresponding to a structure factor
\begin{equation}
\Gamma^\text{(L)}_{\alpha\beta,\gamma\delta}\propto \Pi^{(1)}+\Pi^{(2)}+\Pi^{(6)}=\delta_{\alpha\gamma}\delta_{\beta\delta}.
\end{equation}
This result characterizes the behavior of a \textit{scalar wave}, and coincides with the prediction of the paraxial wave equation (\ref{Paraxial_eq}): the complex polarization vector stays \emph{fixed} during the multiple-scattering process ($\alpha=\gamma$ and $\beta=\delta$), as signaled by the prefactor $|\be\cdot\be'^*|^2$ which varies from 0 to 1 depending on the polarization detection setup. 
We show in the first two rows of Table \ref{table1} the values of $\mathcal{F}_\text{L}=\mathcal{F}_\text{C}(\bk_\perp=-\bk_0)=|\be\cdot\be'^*|^2$ in  four main experimental setups where 
linearly polarized light is analyzed along the parallel ($l\parallel l$) or perpendicular ($l\perp l$)  channels, or where circularly polarized light is analyzed in channels of same ($\sigma\parallel \sigma$)  or opposite ($\sigma\perp \sigma$)  polarization. As intuition suggests, the diffusive ring and the CBS peak are visible in the  $l\parallel l$ and $\sigma \parallel \sigma$ channels only ($\be'=\be$). 

The opposite situation $z\gg z_{1,2}$, on the contrary, \textit{cannot} be described by the paraxial wave equation. In this limit, we find
\begin{equation}
\mathcal{F}(\be,\be',z\gg z_{1,2},\bk_\perp)=\frac{1}{2}\left[1+|\be\cdot\be'|^2e^{-D(\bk_\perp+\bk_0)^2z}\right],
\end{equation}
corresponding to a structure factor
\begin{equation}
\Gamma^\text{(L)}_{\alpha\beta,\gamma\delta}\propto \Pi^{(6)}=\frac{1}{2}\delta_{\alpha\beta}\delta_{\gamma\delta}.
\end{equation}
The values of $\mathcal{F}_\text{L}=1/2$ and $\mathcal{F}_\text{C}(\bk_\perp=-\bk_0)=|\be\cdot\be'|^2/2$ in the four polarization channels are displayed in the last two rows of Table \ref{table1}: the diffusive signal is visible with the same probability in all channels. This indicates that polarization has been \textit{randomized} by the multiple scattering process. The CBS peak, on the other hand, as a high visibility  in the channel $l\parallel l$ and, more unexpectedly, in the channel $\sigma\perp \sigma$ ($\be'=\be^*$): for circularly-polarized incident light the CBS peak is visible only from light of opposite circular polarization. This unusual property is a consequence of the vector nature of light, on which we will come back to in Sec. \ref{RT_Sec} below.
\begin{table}
\begin{tabular}{|>{\centering}p{2.0cm}|>{\centering}p{1.2cm}|>{\centering}p{1.2cm}|>{\centering}p{1.2cm}|>{\centering}p{1.2cm}|}
\hline
channels & $\sigma\parallel \sigma$ & $\sigma\perp \sigma$&$l\parallel l$&$l\perp l$ \tabularnewline
\hline\hline
$\mathcal{F}_\text{L}(z\ll z_{1,2})$&
1&0&1&0
\tabularnewline
\hline
$\mathcal{F}_\text{C}(z\ll z_{1,2})$&
1&0&1&0
\tabularnewline
\hline
$\mathcal{F}_\text{L}(z\gg z_{1,2})$&
1/2&1/2&1/2&1/2
\tabularnewline
\hline
$\mathcal{F}_\text{C}(z\gg z_{1,2})$&
0&1/2&1/2&0
\tabularnewline
\hline
\end{tabular}
\caption{\label{table1}
Values of the diffusive, $\mathcal{F}_\text{L}$ [Eq. (\ref{FL_eq})] and CBS, $\mathcal{F}_\text{C}$ [Eq. (\ref{FC_eq})], contributions of the momentum distribution in the four polarization channels, for $z\ll z_{1,2}$ (scalar regime) and $z\gg z_{1,2}$ (vector regime).}
\end{table}

\begin{figure}
\centering
\includegraphics[width=1\linewidth]{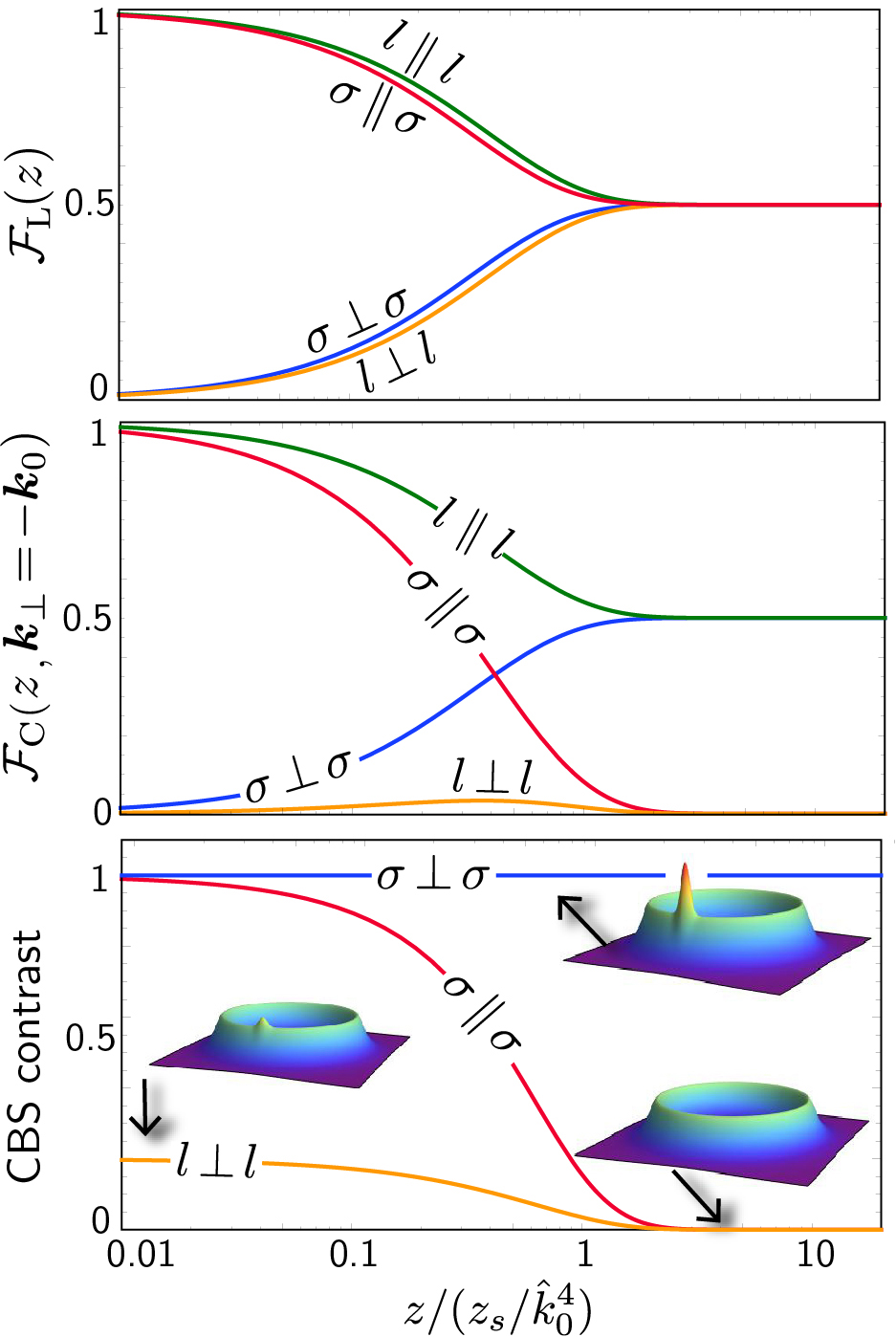}
\caption{
\label{F_plots_fig}
Top panel: Diffusive contribution $\mathcal{F}_\text{L}$ to the momentum distribution as a function of $z$, Eq. (\ref{FL_eq}), in the four polarization channels.
Middle panel : CBS contribution $\mathcal{F}_\text{C}$ at $\bk_\perp=-\bk_0$, Eq. (\ref{FC_eq}),  in the four channels. 
Bottom panel: contrast of the CBS peak,  $\mathcal{F}_\text{C}/\mathcal{F}_\text{L}$ (curves $\sigma\perp\sigma$ and $l\parallel l$ overlap).
The insets display the shape of the momentum distribution in the plane $(k_x,k_y)$, in three configurations where the CBS peak is fully contrasted, partially contrasted and not present. Note that  in the channels $\sigma\perp\sigma$ and $l\perp l$, the CBS peak and the diffusive ring are both very small when $z\ll z_{1,2}$, but their ratio is finite.
}
\end{figure}

For completeness, we plot in Fig. \ref{F_plots_fig} the magnitude of the diffusive and CBS contributions $\mathcal{F}_\text{L}$, Eq. (\ref{FL_eq}), and $\mathcal{F}_\text{C}(\bk_\perp=-\bk_0)$, Eq. (\ref{FC_eq}), as a function of $z$ in the four polarization channels. The ratio  $\mathcal{F}_\text{C}(\bk_\perp=-\bk_0)/\mathcal{F}_\text{L}$ is the contrast of the CBS peak. It is also shown in the figure, together
with the full 2D shape of the momentum distribution (insets).
Note that in the channels $\sigma\perp\sigma$ and $l\perp l$, the CBS peak and the diffusive ring are both very small when $z\ll z_{1,2}$ [they scale as $z/z_{1,2}$, see Eqs. (\ref{FL_eq}) and (\ref{FC_eq})], but their ratio is finite. In particular, in the channel $l\perp l$, the CBS contrast equals $z_2/z_1-1=1/5$.

\subsection{Reciprocity theorem in (2+1) dimensions}
\label{RT_Sec}

In the vector regime $z\gg z_{1,2}$, we have seen that the CBS peak is only visible in channels $l\parallel l$ and, more surprisingly, $\sigma\perp \sigma$. This somewhat counter-intuitive behavior can be understood by general symmetry arguments based on  time-reversal symmetry and homogeneity of the medium along $z$, as we now show.

Consider the wave amplitude $\mathcal{A}^{1\to\text{N}}(\bk_0,\be;\bk_\perp,\be';k_z)$ associated with a multiple scattering sequence $1\to \text{N}$ from the initial state $(\bk_0,\be)$ to the final state $(\bk_\perp,\be')$, with $k_z$ conserved.
The CBS interference is constructed by pairing this amplitude with its counter-propagating partner $\mathcal{A}^{\text{N}\to 1}(\bk_0,\be;\bk_\perp,\be';k_z)$. To find the condition of constructive interference, we first apply the reciprocity theorem pertained to time-reversal symmetry  \cite{Saxon55, Montambaux07}:
\begin{equation}
\label{reciprocity_eq}
\mathcal{A}^{\text{N}\to 1}(\bk_0,\be;\bk_\perp,\be';k_z)=
\mathcal{A}^{1\to\text{N}}(-\bk_\perp,\be'^*;-\bk_0,\be^*;-k_z).
\end{equation}
Owing to the independence of disorder on the longitudinal coordinate $z$, the right-hand-side of Eq. (\ref{reciprocity_eq}) is also parity-symmetric with respect to the variable $k_z$. The CBS interference is therefore constructive provided $\mathcal{A}^{1\to\text{N}}(\bk_0,\be;\bk_\perp,\be';k_z)$ $=\mathcal{A}^{1\to \text{N}}(-\bk_\perp,\be'^*;-\bk_0,\be^*;k_z)$. This equality is obviously satisfied when:
\begin{equation}
\bk_\perp=-\bk_0,\ \be'=\be^*,
\end{equation} 
i.e. the CBS peak is fully contrasted in the channels $l\parallel l$ and $\sigma\perp\sigma$, as found in the previous section.

\subsection{Total distribution and norm conservation}

It is also interesting to evaluate the total momentum distribution, $\overline{|\textbf{E}(\bk_\perp,z)|^2}$, which does not require any particular polarization detection setup. This quantity is readily obtained from Eq. (\ref{momentum_final}) by summing over outgoing polarization vectors $\be'$. Using that $\sum_{\be'}|\be\cdot\be'|^2=\sum_{\be'}|\be\cdot\be'^*|^2=1$, we infer:
\begin{align}
&\overline{|\textbf{E}(\bk_\perp,z)|^2}=\frac{8k}{z_s}\frac{
1
}{(2k/z_s)^2+(\bk_\perp^2-\bk_0^2)^2}\nonumber\\
&\times \left[1+\frac{1}{2}e^{-D(\bk_\perp+\bk_0)^2z}(1-e^{-z/z_1}+2e^{-z/z_2})\right].
\label{momentum_tot_final}
\end{align}
The CBS contrast measured from the total distribution thus varies from 1 in the scalar regime $z\ll z_{1,2}$, to 1/2 in the vector regime  $z\gg z_{1,2}$. These two values are easily understood from the results in Table \ref{table1}: the diffusive ring and the CBS peak are present in the same two polarization channels at small $z$, whereas at large $z$ the ring shows up in all channels and the CBS peak in only half of them. 
Notice, finally, that if we neglect the (small) contribution of CBS, Eq. (\ref{momentum_tot_final}) fulfills, as required, the conservation of normalization for all $z$:
\begin{equation}
\int\frac{d^2\bk_\perp}{(2\pi)^2}\overline{|\textbf{E}(\bk_\perp,z)|^2}=1.
\end{equation}

\subsection{Physical interpretation of $z_{1,2}$}

Let us finally comment on the two characteristic times $z_{1,2}$. They are both on the order of 
\begin{equation}
z_p\equiv\frac{z_s}{\hat{k}_0^4}\sim \frac{z_s}{\theta^4},
\end{equation}
where $\theta$ is the angle made by the incident beam with the optical axis (see Fig. \ref{3Dscheme}).
We have seen above that this time scale separates a scalar regime where polarization is fixed, from a vector regime where polarization is randomized. $z_p$ can therefore be interpreted as the time needed to \textit{randomize the direction of polarization} in (2+1) dimensions. This interpretation can be confirmed by the following qualitative argument. The incident beam, of wave vector $\bk=(\bk_0,k_z)$, has the polarization $\be$. After the first scattering event on a refractive-index fluctuation, the light is scattered into a direction $\hat{\bk}$$^\prime$ and acquires a (unit)
polarization $\be'$ which fulfills \cite{MacKintosh89}
\begin{equation}
\be'=\frac{\be-(\hat{\bk}'\cdot\be)\hat{\bk}'}{\sqrt{1-|\bk'\cdot\be|^2}}.
\label{Deltaepsilondef}
\end{equation}
In the right-hand side,  the dot product $\hat{\bk}'\cdot\be \equiv \bk'\cdot \be_\perp+\epsilon_z \hat{k}_z\sim \hat{k}_0$, since $|\hat{k}_z|\sim |\be_\perp|\sim 1$ and $|\epsilon_z|\sim|\hat{\bk}$$'$$|\sim \hat{k}_0\equiv k_0/k$ for an incident beam almost along $z$. 
By expanding Eq. (\ref{Deltaepsilondef}) to leading order in $\hat{k}_0\ll 1$, we find that 
the 
change in polarization in the plane $(x,y)$, $\Delta\be_\perp\equiv\be'_\perp-\be_\perp$, is 
on the order of
\begin{equation}
|\Delta\be_\perp|\sim
\hat{k}_0^2.
\end{equation}
After a random walk of $N=z/z_s$ scattering events, the polarization subsequently changes by an amount $|\Delta\be_\perp(N)|\sim\sqrt{N}|\Delta\be_\perp|\sim\sqrt{N}\hat{k}_0^2$. This change becomes on the order of 1 when 
\begin{equation}
N\sim\frac{1}{\hat{k}_0^4}\Leftrightarrow z\sim\frac{z_s}{\hat{k}_0^4}\equiv z_p,
\end{equation}
which confirms the physical interpretation of $z_p$ given above. The existence of the two time scales $z_1$ and $z_2$ can be understood from Eq. (\ref{FL_eq}): in the channels of linear polarization, $\mathcal{F}_\text{L}\propto 1\pm e^{-z/z_2}$, while in the channels of circular polarization, $\mathcal{F}_\text{L}\propto 1\pm e^{-z/z_1}$. We conclude that $z_2$ (resp. $z_1$) is the time needed to randomize the polarization of a linearly (resp. circularly) polarized beam. 

We can estimate $z_p$  for parameters used in state-of-the-art experiments. In the recent work \cite{Boguslawski17} for instance, $\lambda\simeq 532$ nm and light is detected at $z\simeq 20$ mm. We estimate the largest $\hat{k}_0$ used in this paper to $\hat{k}_0\simeq 0.17$, and the mean free path $\ell\simeq 15\, \mu$m. This yields $z_s=\ell/\hat{k}_0\simeq 90\, \mu$m, and thus $z_p=z_s/\hat{k}_0^4\simeq 10$ cm. This value is not unreasonably larger than $z$. In view of detecting the cross-over discussed in the present paper, $z_p$ could be decreased by using a slightly larger incident angle or a stronger disorder.

\section{Conclusion}

We have developed a general theory of multiple scattering of light in (2+1) dimensions. 
Starting from the exact Helmholtz equation, we have computed the full vector intensity tensors associated with classical diffusion and with the CBS effect.

Our results demonstrate that in this geometry multiply scattered light experiences a cross-over as it propagates along the effective time axis $z$. This cross-over takes place around a characteristic $z_p$ that corresponds to the time needed for light
to randomize its polarization direction.
Due to the peculiar anisotropic structure of the medium, $z_p$ is much larger than the scattering time (unlike in usual 3D disordered media where they are comparable): it varies with the inverse of the fourth power of the incident beam's angle and is inversely proportional to the disorder strength. $z_p$ thus becomes smaller and smaller as the beam is less and less paraxial, and when disorder fluctuations increase.
As long as $z\ll z_p$, the multiple scattering process is the one of a scalar wave, and is well captured by the paraxial wave equation. When $z$ exceeds $z_p$, light starts to behave as a vector wave and the paraxial equation breaks down. In particular, the CBS interference peak becomes only visible in polarization configurations where light initially linearly (resp. circularly) polarized is detected along the same (resp. the opposite) channel.

Although in this paper we have focused on the optical momentum distribution, our approach is very general and can be applied to  the calculation of various physical observables for light in disordered media of dimension (2+1). Furthermore, as it fully accounts for the coupling between polarization and the spatial structure of the field, our theory could be used to unveil the so-called spin-orbit interactions of light in the presence of disorder, known to show up in heterogeneous media and for non-paraxial beams \cite{Liberman92, Bliokh15}. 
 From an experimental point of view finally, probing light transport over the scalar-to-vector cross-over discussed in this paper would be highly interesting at the onset of Anderson localization, to clarify whether the latter still exists beyond $z_p$.

\section*{Acknowledgments}

The author thanks the Agence Nationale de la Recherche (grant ANR-14-CE26-0032 LOVE) for financial support, and Dominique Delande for illuminating discussions and many advice.

\begin{widetext}
\appendix
\section*{Appendix}
\label{appendix}
In this appendix we provide the full expressions of various tensors, valid for any value of $k_z$ from the paraxial regime ($\hat{k}_z=k_z/k\simeq1$) to the regime of grazing incidence ($\hat{k}_z=k_z/k\simeq0$).

The imaginary part of $\boldsymbol{\Sigma}$, first, follows from Eq. (\ref{Sigma_def_eq}):
\begin{equation}
\text{Im}\Sigma_{ij}(k_z)=-\frac{\gamma k^4}{8}\left[\delta_{ij}(1+\hat{k}_z^2)+\delta_{i z}\delta_{j z}(1-3\hat{k}_z^2)\right].
\label{ImSigma_eq}
\end{equation}
The self-energy tensor is anisotropic, which stems from the symmetry axis $z$ of the problem. 
Note that while the self energy does not depend on $\bk_\perp$, as is expected for a delta-correlated potential in the plane $(x,y)$, it \textit{does} depend on $k_z$ because no average along the longitudinal direction is involved, unlike in the conventional scattering theory of three-dimensional (3D) disordered media. The usual expression of the self energy expected for that case, $\text{Im}\Sigma_{ij}(k_z)=-\delta_{ij}\gamma k^4/6$  \cite{vanTiggelen96}, is recovered by averaging Eq. (\ref{ImSigma_eq}) over $\hat{k}_z^2$. 

Once the self energy is known, the average Green tensor follows from a diagonalization of the Dyson equation (\ref{Dyson_eq}) with respect to polarization indices. This procedure leads to:
\begin{align}
\overline{G}_{ij}(\bk_\perp,k_z)
&=\frac{\delta_{ij}}{k^2-\bk_\perp^2-k_z^2-i\text{Im}\Sigma_1(k_z)}
-\frac{\hat{k}_i\hat{k}_j}{k^2-\bk_\perp^2-k_z^2-i\text{Im}\Sigma_2(k_z)}
\label{AvG_final_eq}\\
&+
\frac{1}{1-\hat{k}_z^2}
\left[\frac{1}{k^2-\bk_\perp^2-k_z^2-i\text{Im}\Sigma_1(k_z)}
-\frac{1}{k^2-\bk_\perp^2-k_z^2-i\text{Im}\Sigma_2(k_z)}\right]
(\delta_{i z}\hat{k}_j\hat{k}_z+
\delta_{j z}\hat{k}_i\hat{k}_z-\delta_{iz}\delta_{jz}-
\hat{k}_i\hat{k}_j
),\nonumber
\end{align}
where $\text{Im}\Sigma_1(k_z)=-(\gamma k^4/8)(1+\hat{k}_z^2)$ and $\text{Im}\Sigma_2(k_z)=-(\gamma k^4/8)(2-3\hat{k}_z^2+3\hat{k}_z^4)$. 
Apart from the Born approximation, Eq. (\ref{AvG_final_eq}) constitutes the exact expression of the average Green tensor in disordered media of  dimension (2+1).
It contains three terms which, in general, must be kept to describe the full multiple scattering problem in this geometry. In particular, working with the exact expression of $\overline{G}_{ij}$ is essential to guarantee the conservation of normalization at all $z$ and to obtain the expressions of the structure factor given in Sec. \ref{sec_results}.

We then provide the expressions of the projectors $\Pi^{(n)}_{\alpha\beta,\gamma\delta}$ on the eigensubspaces of the structure factor $\boldsymbol{\Gamma^\text{(L)}}(k_z,q_z, \bq=0)$, and of their corresponding eigenvalues $\lambda_n$.
In the diffusive regime $q_z z_s\ll1$, we find:
\begin{align}
&\Pi^\text{(1)}_{\alpha\beta,\gamma\delta}=
\frac{1}{2}\left(
\delta_{\alpha\gamma}\delta_{\beta\delta}-
\delta_{\alpha\delta}\delta_{\beta\gamma}-
\delta_{\alpha\gamma}\delta_{\beta z}\delta_{\delta z}+
\delta_{\alpha\delta}\delta_{\beta z}\delta_{\gamma z}+
\delta_{\beta\gamma}\delta_{\alpha z}\delta_{\delta z}-
\delta_{\beta\delta}\delta_{\alpha z}\delta_{\gamma z}
\right)\\
&\Pi^\text{(2)}_{\alpha\beta,\gamma\delta}=
\frac{1}{2}\left(
-\delta_{\alpha\beta}\delta_{\gamma\delta}+
\delta_{\alpha\gamma}\delta_{\beta\delta}+
\delta_{\alpha\delta}\delta_{\beta\gamma}-
\delta_{\alpha\gamma}\delta_{\beta z}\delta_{\delta z}-
\delta_{\alpha\delta}\delta_{\beta z}\delta_{\gamma z}-
\delta_{\beta\gamma}\delta_{\alpha z}\delta_{\delta z}-
\delta_{\beta\delta}\delta_{\alpha z}\delta_{\gamma z}+
\delta_{\alpha\beta}\delta_{\gamma z}\delta_{\delta z}\right.\nonumber\\
&\hspace{1.5cm}\left.+
\delta_{\gamma\delta}\delta_{\alpha z}\delta_{\beta z}+\delta_{\alpha z}\delta_{\beta z}\delta_{\gamma z}\delta_{\delta z}
\right)\\
&\Pi^\text{(3)}_{\alpha\beta,\gamma\delta}=
\frac{1}{2}\left(
\delta_{\alpha\gamma}\delta_{\beta z}\delta_{\delta z}+
\delta_{\alpha\delta}\delta_{\beta z}\delta_{\gamma z}+
\delta_{\beta\gamma}\delta_{\alpha z}\delta_{\delta z}+
\delta_{\beta\delta}\delta_{\alpha z}\delta_{\gamma z}-
4\delta_{\alpha z}\delta_{\beta z}\delta_{\gamma z}\delta_{\delta z}
\right)\\
&\Pi^\text{(4)}_{\alpha\beta,\gamma\delta}=
\frac{1}{2}\left(
\delta_{\alpha\gamma}\delta_{\beta z}\delta_{\delta z}-
\delta_{\alpha\delta}\delta_{\beta z}\delta_{\gamma z}-
\delta_{\beta\gamma}\delta_{\alpha z}\delta_{\delta z}+
\delta_{\beta\delta}\delta_{\alpha z}\delta_{\gamma z}
\right)\\
&\Pi^\text{(5)}_{\alpha\beta,\gamma\delta}=
\frac{(1-\hat{k}_z^2)^2}{3-2\hat{k}_z^2+3\hat{k}_z^4}\delta_{\alpha\beta}\delta_{\gamma \delta}-
\frac{2(1-\hat{k}_z^2)}{3-2\hat{k}_z^2+3\hat{k}_z^4}
\left(\delta_{\alpha\beta}\delta_{\gamma z}\delta_{\delta z}
+\delta_{\gamma\delta}\delta_{\alpha z}\delta_{\beta z}
\right)
+\frac{4}{3-2\hat{k}_z^2+3\hat{k}_z^4}\delta_{\alpha z}\delta_{\beta z}\delta_{\gamma z}\delta_{\delta z}\\
&\Pi^\text{(6)}_{\alpha\beta,\gamma\delta}=
\frac{(1+\hat{k}_z^2)^2}{6-4\hat{k}_z^2+6\hat{k}_z^4}\delta_{\alpha\beta}\delta_{\gamma \delta}+
\frac{(1+\hat{k}_z^2)(1-3\hat{k}_z^2)}{6-4\hat{k}_z^2+6\hat{k}_z^4}
\left(\delta_{\alpha\beta}\delta_{\gamma z}\delta_{\delta z}
+\delta_{\gamma\delta}\delta_{\alpha z}\delta_{\beta z}
\right)
+\frac{(1-3\hat{k}_z^2)^2}{6-4\hat{k}_z^2+6\hat{k}_z^4}\delta_{\alpha z}\delta_{\beta z}\delta_{\gamma z}\delta_{\delta z}.
\end{align}
It is easy to check that $\sum_{n=1}^6\Pi^{(n)}_{\alpha\beta,\gamma\delta}=\delta_{\alpha\gamma}\delta_{\beta\delta}$ and $\Pi^{(n)}_{\alpha\beta,ij}\Pi^{(n')}_{ij,\gamma\delta}=\delta_{nn'}\delta_{\alpha\gamma}\delta_{\beta\delta}$. 
Close to the paraxial regime, all terms of the type $\delta_{iz}$, $i=\alpha,\beta,\gamma$ or $\delta$ are negligible. In particular, $\Pi^{(1)}$, $\Pi^{(2)}$ and $\Pi^{(6)}$ reduce to Eq. (\ref{Pi_expressions_eq}).

The eigenvalues $\lambda_n\equiv \lambda_n(k_z,q_z,\bq=0)$ read, as a function of $\hat{k}_z$ and $q_z$:
\begin{align}
&\lambda_1=\frac{4 \hat{k}_z^2}{3-2\hat{k}_z^2+3\hat{k}_z^4}
-iq_z\frac{64 k_z}{\gamma k^4}\frac{\hat{k}_z^2}{(3-2\hat{k}_z^2+3\hat{k}_z^4)^2}\\
&\lambda_2=\frac{1}{2+2\hat{k}_z^2}+\frac{2\hat{k}_z^2}{3-2\hat{k}_z^2+3\hat{k}_z^4}+\frac{\hat{k}_z^4}{4-6\hat{k}_z^2+6\hat{k}_z^4}
-iq_z\frac{4 k_z }{\gamma k^4}\!\!\left[\frac{1}{(1+\hat{k}_z^2)^2}+\frac{\hat{k}_z^4}{(2-3\hat{k}_z^2+3\hat{k}_z^4)^2}+\frac{8\hat{k}_z^2}{(3-2\hat{k}_z^2+3\hat{k}_z^4)^2}\right]\\
&\lambda_3=-\frac{2}{3}+\frac{2-2\hat{k}_z^2}{3-2\hat{k}_z^2+3 \hat{k}_z^4}+\frac{4}{6-9\hat{k}_z^2+9\hat{k}_z^4}
-i q_z\frac{32 k_z (1-\hat{k}_z^2) }{\gamma k^4}\!\!\left[\frac{\hat{k}_z^2}{2(2-3\hat{k}_z^2+3\hat{k}_z^4)^2}+\frac{1}{(3-2\hat{k}_z^2+3\hat{k}_z^4)^2}\right]
\\
&\lambda_4=\frac{2-2\hat{k}_z^2}{3-2\hat{k}_z^2+3 \hat{k}_z^4}
-i q_z\frac{32 k_z}{\gamma k^4}\frac{1-\hat{k}_z^2}{(3-2\hat{k}_z^2+3 \hat{k}_z^4)^2}\\
&\lambda_5=\frac{2(1-\hat{k}_z^2)^2}{2-\hat{k}_z^2+3\hat{k}_z^6
}-i q_z\frac{16 k_z}{\gamma k^4}\!\!
\left[\frac{1}{2(1+\hat{k}_z^2)^2}-\frac{\hat{k}_z^2}{2(2-3\hat{k}_z^2+3\hat{k}_z^4)^2}+\frac{1}{4-6\hat{k}_z^2+6\hat{k}_z^4}-\frac{1}{3-2\hat{k}_z^2+3\hat{k}_z^4}
\right]
\\
&\lambda_6=1-iq_z\frac{16 k_z}{\gamma k^4}\frac{1}{3-2\hat{k}_z^2+3\hat{k}_z^4}.
\end{align}
The coefficients of the terms in $iq_z$
 define the $\mu_n(k_z)$ factors in Eqs. (\ref{gammaLsum}) and (\ref{gammaCsum}). The eigenvalues $\lambda_2$, $\lambda_3$ and $\lambda_4$ are twice degenerate, all the other are non-degenerate.
We finally provide the factors $\nu_n(k_z)$ in (\ref{gammaCsum})
for the three relevant modes $n=1,2$ and $6$ :
\begin{align}
&\nu_1(k_z)=
\frac{k^2}{(\gamma k^4)^2}\frac{512\hat{k}_z^2(1-\hat{k}_z^2)}{(3-2\hat{k}_z^2+3\hat{k}_z^4)^3}
\\
&\nu_2(k_z)=
\frac{32k^2}{(\gamma k^4)^2}
(1-\hat{k}_z^2)
\left[
\frac{1}{(1+\hat{k}_z^2)^4}
+\frac{\hat{k}_z^4}{(2-3\hat{k}_z^2+3\hat{k}_z^4)^4}
+\frac{32\hat{k}_z^2}{(3-2\hat{k}_z^2+3\hat{k}_z^4)^4}
\right]\\
&\nu_3(k_z)=
\frac{64k^2}{(\gamma k^4)^2}
(1-\hat{k}_z^2)
\left[
\frac{2+\hat{k}_z^2}{8(1+\hat{k}_z^2)^2}
+\frac{3(2-\hat{k}_z^2)}{8(2-3\hat{k}_z^2+3\hat{k}_z^4)^2}
-\frac{3(1-\hat{k}_z^2)}{8(2-3\hat{k}_z^2+3\hat{k}_z^4)}
+\frac{2-3\hat{k}_z^2}{4(3-2\hat{k}_z^2+3\hat{k}_z^4)}\right].
\end{align}
Close to the paraxial regime $\hat{k}_z\simeq 1$, the three coefficients $D_n=\nu_n(k_z)/\mu_n(k_z)$ ($n=1,2$ and $6$) reduce to Eq. (\ref{diff_eq}).\newline

\end{widetext}

\end{document}